\documentclass[12pt,preprint]{aastex}

\shorttitle{Galactic Gamma-Ray Halo of  NGC 253}
\shortauthors{Itoh et al.}

\begin{document}


\title{Galactic Gamma-Ray Halo of \\
the Nearby Starburst Galaxy NGC 253}


\author{C. Itoh\altaffilmark{1} , R. Enomoto\altaffilmark{2},
S. Yanagita\altaffilmark{1}, T. Yoshida\altaffilmark{1},
and T.G.Tsuru\altaffilmark{3}}
\affil{1 Faculty of Science, Ibaraki University, Mito, Ibaraki 310-8512,
Japan; cito@icrr.u-tokyo.ac.jp, yanagita@mx.ibaraki.ac.jp,
yoshidat@mx.ibaraki.ac.jp}
\affil{2 ICRR, University of Tokyo, Kashiwa, Chiba 277-8582, Japan;
enomoto@icrr.u-tokyo.ac.jp}
\affil{3 Department of Physics, Faculty of Science,
Kyoto University, Sakyo-ku, Kyoto 606-8502, Japan;
tsuru@cr.scphys.kyoto-u.ac.jp}








\begin{abstract}
Recently, the CANGAROO-II telescope detected diffuse
TeV gamma-ray emission from a nearby
edge-on starburst galaxy, NGC~253.
The emission mechanism  is discussed in this report.
We review the emissions of radio-to-TeV gamma-rays
from NGC~253, and present a model
of the non-thermal emissions
due to synchrotron radiations and inverse Compton scatterings.
A halo model  successfully explains
the multiband spectrum of NGC~253.
\end{abstract}


\keywords{cosmic rays---galaxies: halos---
galaxies: individual(NGC 253)---galaxies: starburst
---gamma rays: theory}


\section{Introduction}
The CANGAROO-II imaging atmospheric Cherenkov telescope
has detected TeV gamma-ray emission
from a nearby edge-on starburst galaxy, NGC~253
\citep{Itoh2002a, Itoh2002b}.
This TeV emission is spatially extended and temporally steady,
whose nature is different from that of previously observed
extragalactic objects of the active
galactic nuclei (AGN) class.
This is the first detection of TeV gamma-rays from a normal-sized
spiral galaxy like our Galaxy.
We can learn how high-energy particles propagate on the galactic scale
from the observations of NGC~253 by TeV gamma-rays.
\par

The acceleration and propagation of galactic cosmic rays (GCRs),
which are observed directly near
the earth, are among the big topics in physics and astrophysics.
The distribution of GCR with an energy of
less than $\sim$10 GeV in the
Galaxy has been surveyed by tracing diffuse
gamma-rays which are produced by the interaction of
GCR with the interstellar
medium\citep{Hunter1997}.
The resultant distribution of diffuse gamma-rays as a function of the galactic
longitudes and latitudes has been argued based on the
scenario of the SNR origin of GCR.
In the TeV energy range \citep{LeBohec2000, Aharonian2001},
searches for
diffuse
gamma-rays in our Galaxy have been also carried out to obtain the spatial
distribution of TeV cosmic rays.
It is, however, difficult
due to the limitations of the present techniques.
In some respects, studies of nearby galaxies which are near
enough to obtain detectable fluxes from outside are more reasonable than
that of our Galaxy from inside.
A further question naturally arises as to whether other galaxies also
harbour cosmic-rays like ours.
Photons from radio to gamma-rays produced by cosmic-rays would
reveal their existence.
The electrons of GeV energies associated with edge-on galaxies
have been known by radio
observations \citep{Berzinskii1990, Beck1997}.
EGRET demonstrated clear evidence for the existence of
GeV particles in LMC \citep{Sreekumar1992}; however,
no evidence for such high-energy particles as TeV
in normal galaxies had been known, except for that from our Galaxy.
Obviously, investigating nearby galaxies like NGC~253 is a key.
\par

In Section 2  we review the emissions of radio-to-TeV
gamma-rays of this galaxy.
In Section 3 we present a model which explains the observed
multiband spectrum and the spatial distribution of cosmic rays in
NGC~253.
We discuss the propagation of high-energy particles into the halo,
the reacceleration of
particles due to the galactic wind, and the total cosmic-ray energy of NGC~253
in Section 4.
\par

\section{Multi-wavelength Spectrum}
The distance to NGC~253 is as close as
2.5 Mpc \citep{Vaucouleurs1978}
with a visual size of 0.3$^{\circ}$
from Earth.
This is one of the best targets where the cosmic-rays density is
expected to be high
due to its high supernova rate \citep{Voelk1996}.
Here, we summarize previous studies at NGC~253 in various photon
energies. The compiled fluxes and
upper limits on NGC~253 are shown in Fig. \ref{fig1}.
\par

The EGRET instrument on board the Compton Gamma Ray
Observatory(CGRO)
has reported stringent upper limits for
GeV gamma-rays \citep{Sreekumar1994, Blom1999}.
On the other hand, the OSSE on board CGRO has claimed
the detection of 50$\sim$200 keV gamma-rays \citep{Bhattacharya1994}.
The PDS on board BeppoSAX gives a contradicting upper limit
on the hard X-ray emission just below OSSE's energy band
\citep{Cappi1999}.
\par

At TeV energies, CANGAROO-II has detected diffuse emission.
The size of the emitting region is similar
to, or larger than, the optical image of NGC~253
\citep{Itoh2002a, Itoh2002b}.
A simple power-law fit gave
a soft spectral photon index of $3.74\pm0.27$\citep{Itoh2002b},
suggesting a cutoff.
\par

The radio continuum emission consists of three main components: a
central region, a disk component, and a halo extending
$\sim 9~{\rm kpc}$ {\citep{Hummel1984,Carilli1992}}.
The radio spectral energy index is $-0.7\pm 0.1$ in the disk,
and it is steepened to
$\simeq -1$ in the halo. The fractional
polarization increases with the distance from the plane (less than
4\% in the disk, up to 15\% in the halo in 1.4GHz
{\citep{Carilli1992}}, and reaching 40\% at 10GHz {\citep{Beck1994}}).
The
turbulent magnetic fields estimated from the fractional polarization are
$\sim 17\mu{\rm G}$ in the disk and $\sim6\mu{\rm G}$ in the
halo {\citep{Beck1994}}.
The radio emission in the central region of $\sim 150\times 70 {\rm
pc}$ shows a flat spectrum with an energy index of $\sim -0.43$
{\citep{Hummel1984}. From high-resolution radio
observations of the inner 200 pc of NGC~253,
\citet{Ulvestad1997} detected $\sim 64$ compact
radio sources, half of which are in the
HII regions with flat spectra.
The other half are presumably synchrotron supernova
remnants.

The X-ray structure of NGC~253 is complex, and consists of point
sources, hot intersteller matter (ISM) in the disk and the central region,
and the halo.
The diffuse X-ray emission from the halo was first detected with
Einstein, and followed up by ROSAT, ASCA, BeppoSAX, XMM-Newton, and
Chandra {\citep{Watson1984, Fabbiano1988, Pietsch2000, Ptak1997,
Pietsch2001, Strickland2000b, Strickland2002, Cappi1999, Weaver2002}}.
X-ray spectral studies of the disk and the halo
show the existence of optically thin thermal
plasmas at temperatures of $kT = 0.1\sim 0.4$~keV.
Chandra and XMM-Newton observations of the central region of NGC~253
reveal the existence of a heavily absorbed high-temperature
thermal plasma component ($kT \sim 6$~keV)
with an ionized iron K emission line
{\citep{Pietsch2001, Weaver2002}}.
This absorbed component is also seen in
the spatially integrated spectra over the whole galaxy obtained
with the previous X-ray satellites {\citep{Ptak1997, Cappi1999, Ohashi1990}}.
Based on a magnetic field strength of $270~\mu{\rm G}$
in the central region, estimated from the standard minimum energy assumption,
\citet{Weaver2002} suggested that the inverse Compton process
makes only a minor contribution to the hard X-ray emission.
\par




\section{Inverse Compton Emission Halo Model}

We first discuss the emission mechanism at  gamma-ray energies
from the multi-wavelength spectrum, as shown in Fig. \ref{fig1}.
\cite{Goldshmidt1995} reported that
the gamma-ray emission detected by OSSE (the blank circles
in Fig. \ref{fig1})
was attributed to the Inverse Compton Scattering (IC)
of ubiquitous far-infrared(FIR) photons from dusts observed around
the center of the galaxy.
However, a simple extrapolation to the GeV range
exceeds the 2 $\sigma$ upper limit of $3.4\times10^{-8}$
~cm$^{-2}$ s$^{-1}$ \citep{Blom1999} obtained
by EGRET (the red triangles).
The multi-wavelength spectrum suggests that the origin of TeV gamma-rays
should be different from that of sub-MeV gamma-rays.
\par

If we assume that the energy
spectrum of particles is a power law with an exponential cut,
and that the power-law index is softer than $-2$,
the index of the differential photon
spectra in GeV energies due either to the emissions
of $\pi^{0}$ decay produced
by the collisions of high-energy protons with the matter
or non-thermal
bremsstrahlung of the high-energy electrons
also becomes softer than $-2$, leading to a contradiction with the
EGRET upper limit.
\par
The remaining possibility of the emission mechanism of GeV and TeV gamma-rays
is IC scattering.
When the power-law index of the electron energy spectrum is assumed
to be $\sim -2$ ,
the index of the differential photon spectrum of the IC
becomes nearly $-1.5$.
\par

The very large radio halo of $\sim 9~{\rm kpc}$ \citep{Carilli1992} would
suggest the existence of a population of very high-energy electrons
which emit our TeV gamma-rays other than those concentrated near to the
central or disk region of the galaxy, which might emit photons observed by
OSSE.
On the other hand, there are sure target photons for IC, such as CMB
as well as very abundant FIR photons.
Here, we quantitatively examine whether
these electrons and photons are responsible for TeV emissions.

We assume that the total number spectrum
of electrons at the source has the following form:
\begin{equation}
N_{e}(\gamma)=N_{e0} \gamma^{-p} \exp(-\gamma/\gamma_{m}),
\end{equation}
where $N_{e0}$ is an electron density factor, $\gamma$
the Lorentz factor, and $\gamma_{m}$
the maximum Lorentz factor of the electrons (a cutoff), respectively.
The spectrum of the IC emission is given by
\begin{equation}
\frac{dF}{dE}=\frac{1}{4 \pi D^{2}}\int \int N_{e}(E)
n_{ph}(\epsilon)c\sigma(E_{\gamma},\epsilon,E)d \epsilon dE,
\end{equation}
where $D$ is the distance to the source,
$n_{ph}$ the number spectrum of the soft photon,
$c$ the speed of  light, $\epsilon$ the energy of the target photons,
$E_{\gamma}$ the energy of the scattered photons,
$E$ the energy of the electrons,
and $\sigma(E_{\gamma},\epsilon,E)$
the Klein-Nishina cross section for IC scattering,
respectively.
\par

We take the FIR photon spectrum model,
in which the dust absorption efficiency is proportional
to the photon frequency, as described in
\citet{Goldshmidt1995}.  The total FIR luminosity is
$L_{\rm IR}=1.97\times10^{10}L_{\odot}$, with
$1.45\times10^{10}L_{\odot}$
from cool dust having a temperature $T_{c}=36.8$ K and
$0.52\times10^{10}L_{\odot}$
from warm dust with $T_{w}=172.8$ K, respectively.
Here, we assume that the photon field is isotropic.
The energy density
of the photon field $U_{\rm IR}$ at radius $R_{s}$ is
approximated as $U_{\rm IR}=L_{\rm IR}/(\pi R_{s}^{2}c)$.
We simply adopt  the average value of $U_{\rm IR}$ in a spherical region
with a size of $R_{s}$  as the photon density.
The average photon density ($<U_{\rm IR}>$) is
$\int_{0}^{R_{s}} U_{\rm IR}(r) 4 \pi  r^2 dr/(4 \pi {R_{s}}^3/3)=
3U_{\rm IR}(R_{s})$.
The optical photon field is assumed to be
a diluted blackbody with a temperature of $6000$ K
and a luminosity of $L_{\rm B}=2.65\times10^{10}L_{\odot}$
\citep{Pence1980}.
Assuming the value of the electron spectrum index $p$,
we could determine the two parameters  $N_{e0}/(4 \pi D^{2})$
and $\gamma_{m}$ for the electron spectrum as fitting the CANGAROO-II
flux and keeping the EGRET upper limit.
\par

Here, we adopt a one-zone model with two different sizes
for simplicity.
First, we adopt $R_{s}=10$ kpc, suggested by
the size of TeV emitting  region \citep{Itoh2002a},
with assuming a cosmic-ray halo.
We considered three populations of photons as the target
of the IC process, IR, optical, and CMB.
The energy densities for each population are
$<U_{\rm IR}>=1.6$ eV cm$^{-3}$,
$<U_{\rm opt}>=2.2$ eV cm$^{-3}$,
and $U_{\rm CMB}=0.26$ eV cm$^{-3}$, respectively.
Although $<U_{\rm opt}>$ is greater than $<U_{\rm IR}>$,
its contribution was suppressed by the Klein-Nishina effect.
In Fig. \ref{fig1},
the solid and dashed lines represent the synchrotron and IC
emission models
with the power-law index of the electrons
being $p=2$ and $2.2$, respectively.
Here, the value of $p$ was
chosen based on constraints by the shock acceleration ($p\geq 2$)
and reconciliation between the EGRET upper limits and
the CANGAROO-II TeV results ($p=2.4$) \citep{Itoh2002a, Itoh2002b}.
The fluxes of the TeV gamma-rays and the upper
limit of EGRET restricted the parameter choices of the electron
spectrum:
$N_{e0}/(4 \pi D^{2})=6.8\times10^{8}$ cm$^{-2}$
and $\gamma_{m}=2.5\times10^{6}(1.3$TeV)
for $p=2 $,
$N_{e0}/(4 \pi D^{2})=8.4\times10^{9}$ cm$^{-2}$
and $\gamma_{m}=3.6\times10^{6}(1.9$TeV)
for $p=2.2$,
and
$N_{e0}/(4 \pi D^{2})=8.5\times10^{10}$ cm$^{-2}$
and $\gamma_{m}=4.7\times10^{6}(2.4$TeV)
for $p=2.4$,
 respectively.
The IC fluxes at $\sim 2$ keV for $p=2.2 \sim 2.4$ are predicted to
be the same as the Chandra X-ray data in the halo region. This X-ray
data imply these halo models favor the value of $p <2.2 \sim 2.4$.
We find the total energies and average energy densities of the electrons
to be within $R_{s}=10$ kpc:
$5.9 \times 10^{54}$ erg and $0.03$ eV cm $^{-3}$ for $p=2$,
$2.4\times10^{55}$ erg and $0.12$ eV cm $^{-3}$ for $p=2.2$,
and
$1.3\times10^{56}$ erg and $0.66$ eV cm $^{-3}$ for $p=2.4$,
respectively.
The strength of the magnetic field ($B$) were estimated to be
$2.5\mu{\rm G}$ for $p=2$,
$1.7\mu{\rm G}$ for $p=2.2$,
and $1.3\mu{\rm G}$ for $p=2.4$, respectively,
by adjusting the
calculated  synchrotron emissions
to the radio halo data. These values are smaller than the estimations
obtained by the fractional polarization in radio measurements
\citep{Beck1994}.
\par

On the other hand,  we examined the case for a disk component of cosmic-rays.
We  adopted a higher average photon density within
$R_{s}=3$ kpc.
The values of the average photon energy densities are
$<{U}_{\rm IR}>=18$ eV cm$^{-3}$ and
$<{U}_{\rm opt}>=24$ eV cm$^{-3}$.
In Fig. \ref{fig1}, the dotted lines represent the
emission models from a disk with the power-law index
of the electrons of $p=2.4$, which  was estimated
based on the radio data of the disk.
The parameters of the electron spectrum are
$N_{e0}/(4 \pi D^{2})=7.7\times10^{9}$ cm$^{-2}$
and $\gamma_{m}=4.7\times10^{6}(2.4$ TeV).
We found the total energy and
average energy density of the electrons
within $R_{s}=3$ kpc to be:
$1.2 \times 10^{55}$ erg and $2.2$ eV cm $^{-3}$.
The strength of the magnetic field, $B=8.2\mu{\rm G}$,
reproduces the radio fluxes from the disk.
 In the case of a soft electron spectrum with $p=2.4$,
fittings for the CANGAROO-II fluxes and the upper limits
of EGRET
are rather difficult, as shown in Fig.~\ref{fig1}.
The high-energy electron halo model explained better the radio and GeV-TeV
emission.
\par

The halo model, however, cannot describe any emissions in the
keV-MeV energy region.
\citet{Goldshmidt1995}  introduced the IC model with a very localized FIR
and high-energy electrons in order to explain the OSSE result.
Their IC model should require a cutoff in the energy
spectrum of electrons to reconcile with the EGRET upper limits.
We could reproduce the OSSE flux by adopting an
average photon density of
$<U_{\rm IR}>=1.8\times10^3$ eV cm$^{-3}$
within $R_{s}=0.3$ kpc,
and taking the following parameters of the electron spectrum:
for example, in case of $p=2.3$, $N_{e0}/(4 \pi D^{2})=4.5\times10^{9}$
cm$^{-2}$
and $\gamma_{m}=1.2\times10^{4}(6.2$ GeV).
The total energy and average energy density of the electrons are
$8.5 \times 10^{54}$ erg and $1.6\times10^{3}$ eV cm $^{-3}$,
respectively.
Such a high density and low cutoff energy may be reasonable,
considering the starburst characteristics and
FIR density around the center of this
galaxy.
This is also consistent with the EGRET upper limits, as shown
 in Fig. \ref{fig1} (the long-dashed line).
The synchrotron emission was also reproduced along
with the strengths of the magnetic field, $B=9.6\mu{\rm G}$,
which is almost the same value as that adopted by \citet{Goldshmidt1995}.
However, the model IC flux at or above $\sim 2$ keV is more than one order
of magnitude higher than the Chandra and XMM X-ray data, which were
corrected for the photoelectric absorption in both the Galaxy and
NGC 253. The model IC flux in hard X-ray energy band is also higher than
the BeppoSAX PDS X-ray data, while these data were not corrected
for the absorption which was estimated to be negligibly small in the
higher energy. We concluded that some modifications are necessary
for this model.
\par


Hitherto, we considered only the possibility of a leptonic
origin of the observed TeV gamma rays; however, we should
check before reaching any final conclusion whether hadronic models
can explain CANGAROO-II's positive detection consistently
with the EGRET upper limits. In this estimation of gamma-ray
emission by $\pi^o$-decay\citep{Mori1997},
we assume the proton-to-electron ratio
to be 300, which is rather high compared to that of GCRs
($\sim 100$) and accordingly is a favourable case for hadronic models.
We assume the parameters for electrons to be $p=2$ with a higher cutoff
energy of 10 TeV and $N_{e0}/4\pi D^2=6.8\times 10^8~{\rm cm^{-2}}$,
which we have
shown previously.
We also assumed the total mass of the gas which interacts
with protons to be $3 \times 10^9 M_\odot$
\citep{Puche1991, Houghton1997}.
The expected
gamma-ray flux is shown in Fig. \ref{fig1} as the dot-dashed line.
It can be clearly seen that hadronic models fail to explain
the observations, even in the favourable case.

\section{Propagation and Acceleration} 

The high-energy
electrons in the halo
of the galaxy explain most of multiband spectrum of
NGC~253 quite well.
Here, we discuss the origin of these cosmic-ray electrons
and the total cosmic-ray energy,
which should be compared with our Galaxy.
\par

The biggest question is whether high-energy electrons can propagate
out to a distance of $\sim$ 10kpc from the disk of
the galaxy without any severe energy losses.
Electrons may diffuse out to a distance of
$R_{L}\sim2(\kappa t_{cool})^{1/2 }$
within their cooling time ($t_{cool }$) due
to synchrotron and IC losses, where $t_{cool }$ is
inversely
proportional to a
sum of the energy densities of the magnetic field
and the photons in ambient
space, and is also inversely proportional to
the energy of the electrons.
When the diffusion coefficient of electrons
($\kappa$)
of order $3\times10^{29}(E/{\rm GeV})^{0.6}
\rm{cm}^{2} s^{-1}$ is used,
we found that electrons of $1$ TeV
which emit sub-TeV gamma-rays can propagate
to a distance of $\sim9$ kpc,
assuming the strength of magnetic field
to be $2.5 \mu{\rm G}$, and the energy density of FIR photon to be
$U_{\rm IR}(10$ kpc$)$.
The assumption regarding the energy dependence of $\kappa$ is crucial
for this estimation.
\par

The diffuse X-ray emission extending from the nuclear region toward
the halo is thought to be due to the galactic wind (super-wind),
powered by supernovae occurring in the nuclear starburst region
{\citep{Chevalier1985}}, which has been successfully described by many
papers based on hydrodynamical simulations {\citep{Tomisaka1988,
Strickland2000a, Suchkov1994}}.  Continuous supernova explosions heat
up the ISM
in the nuclear starburst region, and creates a bubble of
high-temperature plasmas with $T\sim 10^8{\rm K}$. The bubble expands
while sweeping and pushing cool ISM and ambient gas up towards the
halo. A galactic wind running freely along the galactic minor axis
is formed around $10^7{\rm yr}$ after the start of the starburst
activity. Wind with a velocity of $2000\sim 3000{\rm km\
sec^{-1}}$ finally hits the ambient cool gas and forms a shock in the
halo. This picture explains that
the image of the synchrotron radio halo at $0.33{\rm GHz}$
generally matches the diffuse X-ray emission
{\citep{Pietsch2000}}.
This shock in the halo may accelerate particles by the Fermi process.


The maximum energy of electrons accelerated by this process is estimated
to be 8 $\sim$ 25~TeV if we assume a shock velocity of
$\sim 2000{\rm km\ sec^{-1}}$,
a magnetic field of 2~$\mu$G, and a starburst age of $\sim~10^{7}$~yr by
equating the acceleration rate with the cooling rate due to the
synchrotron and IC losses.
The estimated maximum energy of electrons agrees well with that determined
from TeV gamma-ray spectrum by CANGAROO observation.
Another possibility of particle acceleration in the termination shock of
the galactic wind has been proposed by Jokipii \& Morfill (1985).
These two cosmic-ray acceleration processes may be related
to the rather hard spectrum required for the electrons in the halo.

Finally, we discuss the total cosmic-ray energy in the halo.  The
electron energy density was obtained to be $0.03\sim0.12$ eV cm$^{-3}$,
which is one order higher than that of our Galaxy. If we  assume the proton
to electron ratio to be 100, this implies that the total energy of cosmic
rays in NGC~253 may amount to $5.9\times10^{56} \sim
2.4\times10^{57}{\rm ergs}$, which is one hundred-times higher than
that for our Galaxy. Note that the value for our Galaxy was derived
assuming a disk population of cosmic-rays. If our Galaxy has a halo
with the same cosmic-ray density, the deviation becomes smaller.
The starburst age and the supernova rate in NGC~253 were estimated to
be $\sim 10^7{\rm yr}$ and $\sim 0.3{\rm yr^{-1}}$, respectively
{\citep{Rieke1988}}. Thus, taking the ambiguities (e.g. the proton to
electron ratio) into account, the starburst activity characterizing
NGC~253 may explain the large amount of total energy of cosmic rays.



\acknowledgments
We gratefully acknowledge the CANGAROO-II team for discussions
and suggestions, and thank Dr. M. Mori for his support in
providing us with his code for calculating the gamma-ray
spectrum from the pion decay. We also thank Prof. W.Pietsch
for his kindly providing the information on the XMM spectrum.

\clearpage


\begin{figure}
\plotone{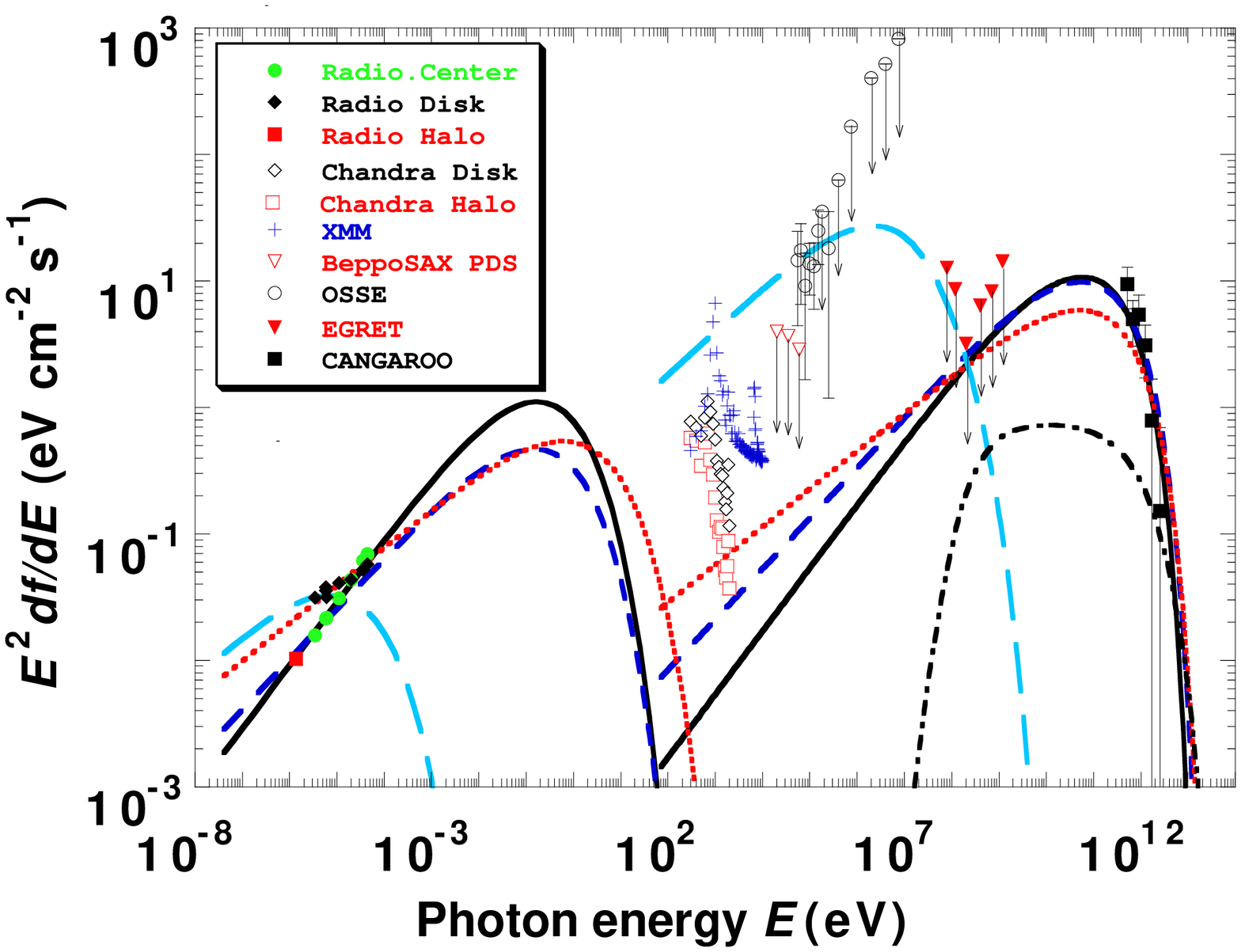}
\caption{
Multi-band spectrum of NGC~253. The black squares were
obtained by CANGAROO-II.
The X-ray data were corrected for photo-absorption in both
the Galaxy and NGC~253,
except that no correction was applied to the
BeppoSAX hard X-ray data.
We note that the
BeppoSAX, OSSE, and EGRET data were not able to spatially resolve
NGC~253, in contrast to the other data.
The lines shown were obtained by estimations described in the text.
}
\label{fig1}
\end{figure}

\end{document}